\documentclass{nature}
\usepackage{graphicx,makecell,multirow}
\usepackage{amsmath,amssymb,color}
\usepackage{lscape}
\usepackage{hyperref}
\usepackage{xcolor}
\usepackage{comment}
\usepackage{ulem}
\hypersetup{colorlinks,allcolors=blue}

\def\arcmin{\hbox{$^\prime$}}
\def\arcsec{\hbox{$^{\prime\prime}$}}
\def\arcdeg{\hbox{$^\circ$}}

\def\farcs{\hbox{$.\!\!^{\prime\prime}$}}
\def\farcss{\hbox{$.\!\!^{\rm s}$}}

\graphicspath{{./}{figures/}}

\title
{Large gas inflow driven by a matured galactic bar in the early Universe}

\begin{document}
\maketitle
\author{Shuo Huang,$^{1,2}$ Ryohei Kawabe,$^{1,3}$ Hideki Umehata,$^{4,2}$ Kotaro Kohno,$^{5,6}$ Yoichi Tamura,$^{2}$ Toshiki Saito$^{7}$}
\begin{affiliations}
\item {National Astronomical Observatory of Japan, Osawa 2-21-1, Mitaka, Tokyo 181-8588, Japan}
\item {Department of Physics, Graduate School of Science, Nagoya University, Furocho, Chikusa, Nagoya 464-8602, Japan}
\item {Department of Astronomy, School of Science, The Graduate University for Advanced Studies (SOKENDAI), Osawa, Mitaka, Tokyo, 181-8588, Japan}
\item {Institute for Advanced Research, Nagoya University, Furocho, Chikusa, Nagoya 464-8602, Japan}
\item {Institute of Astronomy, Graduate School of Science, The University of Tokyo,  2-21-1 Osawa, Mitaka, Tokyo 181-0015, Japan}
\item {Research Center for the Early Universe, Graduate School of Science, The University of Tokyo, 7-3-1 Hongo, Bunkyo-ku, Tokyo, 113-0033, Japan}
\item {Faculty of Global Interdisciplinary Science and Innovation, Shizuoka University, 836 Ohya, Suruga-ku, Shizuoka 422-8529, Japan}

\end{affiliations}

%
%

\begin{abstract}
  Bar structures are present in about half of local disk galaxies\cite{2011MNRAS.411.2026M} and play pivotal roles in secular galaxy evolution. Bars impose a non-axisymmetric perturbation to the rotating disk and transport gas inward to feed central starburst and, possibly, the activity of the nuclear supermassive black hole\cite{2016MNRAS.463.1074C}. They are believed to be long-lived structures\cite{2015A&A...584A..90G,2020MNRAS.491.2547R} and are now identified at redshift $z>2$\cite{2023Natur.623..499C,2023ApJ...945L..10G}. Yet, little is known about the onset and effect of bars in the early cosmic epoch because spectroscopy of distant bars at sufficient resolution is prohibitively expensive. Here, we report a kinematic study of a galactic bar at redshift 2.467, 2.6 billion years after the Big Bang. We observe the carbon monoxide and atomic carbon emission lines of the dusty star-forming galaxy J0107a and find the bar of J0107a has gas distribution and motion in a pattern identical to local bars\cite{2007PASJ...59..117K,2005MNRAS.364..773F,2022ApJ...939...40L}. At the same time, the bar drives large-scale non-circular motions that dominate over disk rotation, funneling molecular gas into its center at a rate of $\approx600$ solar masses per year. Our results show that bar-driven dynamical processes and secular evolution were already at play 11.1 billion years ago, powering active star formation amid the gas-rich and far-infrared luminous growth phase in a massive disk galaxy.
\end{abstract}
J0107a\cite{2014ApJ...781L..39T,2021ApJ...917...94M} is the most massive barred spiral known at $z>2$, and even heavier than the most massive disks in local universe\cite{2016ApJ...817..109O}, with a stellar mass of $4.5^{+1.7}_{-1.4}\times10^{11}M_\odot$, star formation rate of $499^{+357}_{-179}M_\odot$ yr$^{-1}$ from spectral energy distribution modeling of Chandra, JWST NIRCam/MIRI, and Atacama Large Millimeter/submillimeter Array (ALMA) dust continuum data\cite{2023ApJ...958L..26H}. The CO(1\textendash0) emission line detected by the Karl G. Jansky Very Large Array (VLA) indicates a molecular gas mass of $(3.3\pm0.6)\times10^{11}M_\odot$, typical of star-forming galaxies at this redshift and stellar mass\cite{2018ApJ...853..179T}. The rest-frame near-infrared image shows a prominent stellar bar with projected diameter $\approx15$ kpc and two nearly symmetric, pronounced spiral arms starting from the bar ends (Figure 1a). J0107a's face-on (see method) barred spiral morphology, extraordinary brightness powered by its mass, and the largest angular size among $z>2$ bars\cite{2024arXiv240906100G} together provide a unique opportunity for the first clear view of a bar in the formative epoch\cite{2015A&A...584A..90G} without the assistance of gravitational lensing magnification and the resultant delensing uncertainties, which is otherwise extremely challenging at such redshift.

\par To unveil the internal gas distribution and dynamics of J0107a, we have carried out observations of the CO(4\textendash3) and [CI](1\textendash0) emission lines using ALMA. The CO(4\textendash3)/[CI](1\textendash0) and 2.1 mm continuum data, tracing the cold interstellar medium at $\sim0\farcs16$ ($\sim1.5$ kpc at $z=2.467$) spatial resolution, clearly show a gaseous bar with two straight dust lanes on the leading side of the stellar bar (Figure 1b\textendash d), exactly like local strongly barred galaxies, which are the consequence of periodic orbits crowding in a bar-perturbed disk potential and inducing shock\cite{1992MNRAS.259..345A}. The molecular gas is concentrated in the bar region, especially on the west offset ridge as seen in both CO(4\textendash3) and [CI](1\textendash0) (Figure 1c/d). 
The large coherent bar structure seen not only in stellar light but also in molecular gas and dust indicates that the gas disk should have persisted for several dynamical timescales to respond to the gravitational torque of the bar (a dynamical timescale $\sim$ the time for one rotation $\approx200$ Myr in J0107a, see also the following text), which is also comparable to the formation timescale of J0107a $\sim M_\star/$SFR$\sim1$ Gyr. In other words, the bar formed by $z\sim3-4$.
\par
Using the rest-frame near-infrared image and [CI](1\textendash0) emission as proxies to the mass distribution, we compute the gravitational potential and torque field of J0107a  (Figure 2a\textendash c, see methods for details). The torque field shows an apparent four-quadrant shape, again characteristic of a bar-perturbed disk. In stellar mass distribution, bar strength $A_2$ (the ratio between amplitudes of $m=2$ and $m=0$ Fourier modes) reaches 0.8, and the ratio between maximum tangential and radial forces exceeds 0.4 (Extended Data Figure 6), so the bar perturbation is comparable to local strongly barred galaxies\cite{2016A&A...587A.160D}. The azimuthal average of torque on molecular gas is overall negative so the gas loses angular momentum (Figure 2d) and flows inward at a rate of $579_{-174}^{+188}M_\odot$ yr$^{-1}$, more than an order of magnitude higher than $0.01-50M_\odot$ yr$^{-1}$ observed in local galaxies\cite{2009ApJ...692.1623H}. The 888 $\mu$m continuum emission has distribution follows the bar\cite{2023ApJ...958L..26H}, indicating the total SFR of $\approx500M_\odot$ yr$^{-1}$ is confined within the bar region. The gas inflow rate is enough to feed this SFR and might also be responsible for the nuclear supermassive black hole activity (bolometric luminosity $\approx7\times10^{11}L_\odot$\cite{2014ApJ...781L..39T}) in J0107a. 

\par
With the new CO(4\textendash3) data, we characterize J0107a's gas dynamics as a combination of ordered rotation and non-circular motion driven by the bar. In the position-to-velocity (PV) diagram along the major axis, the outermost contour can be described with flat rotation connected to a steeply rising central part (Figure 3b, the dotted line). The widespread low-velocity components at radius $1.5<|r|<5.0$ kpc and velocity $-20<v<40$ km s$^{-1}$, denoted by arrows in Figure 3b, reflect the inflowing gas concentration on bar offset ridge\cite{2009PASJ...61..441H} and dominate over the disk rotation component in intensity.
Notably, the central $|r|<0.8$ kpc region shows two peaks at $|v|>0$ km s$^{-1}$ (the $6\sigma$ contours inside vertical lines in Figure 3b), also seen as the double-peaked spectrum of the center (Figure 3b, the left inserted panel). Because the galaxy is viewed almost face-on (see method), this feature means off-center local maxima in gas distribution and can be interpreted as a nuclear ring, which is a phenomenon often observed within the central 1 kpc radius of nearby barred galaxies\cite{2024MNRAS.528.3613E}.
A high-velocity tail is visible at radius $\sim5$ kpc and extends to line-of-sight (LOS) velocity $\approx-200$ km s$^{-1}$, marked by blue arrows in Figure 3a/b. This component is associated with the southeast spiral arm.
In addition, velocity gradient is present along the minor axis (Figure 3b, the right inserted panel), being blueshifted on the north (the near side, see method). This is the right sign of an outflow, but the absolute velocity of $<100$ km s$^{-1}$ is lower than several hundred km s$^{-1}$ seen in the molecular outflow of high-redshift dusty star-forming galaxies\cite{2020ApJ...905...85S}. Alternatively, the spatial distribution suggests non-circular motion as its origin because its position corresponds to the west base of the bar and a gas cloud north to the core. Such a cloud is not present on the south side, so the minor axis PV diagram appears asymmetric. Then, a plausible explanation for the observed velocity is that the gas flows from pericenter to apocenter along elongated orbits in bar potential\cite{1992MNRAS.259..345A} (Figure 4) and appears to move outward.

\par
In the LOS velocity map  (Figure 3c), the zero-velocity contour is twisted into S-shape, again a common feature of barred galaxies\cite{2005MNRAS.364..773F,2022ApJ...939...40L}. This is caused by non-circular motion superimposing on the disk rotation (Figure 3d, see also methods for details about modeling). The non-circular motion at radii $2<r<4$ kpc dominates over rotation, so the S-shape appears in the LOS velocity map before subtracting rotation. The radius range where non-circular velocity is high overlaps with strong bar perturbation in the gravitational potential (Extended Data Figure 6), highlighting the effect of the bar. 
From the LOS velocity map, we estimate the pattern speed of the bar (i.e., the angular speed at which the bar rotates as a rigid body) to be $67.5\pm17.2$ km s$^{-1}$ kpc$^{-1}$. The ratio between the corotation radius where the bar reaches the circular velocity of the disk and bar semi-major axis length is $R_{\rm CR}/R_{\rm bar}\approx1.3$, where $R_{\rm bar}$ is measured from ellipse isophote fitting of rest-frame near-infrared image\cite{2023ApJ...958L..26H}. This places J0107a in the fast bar regime ($1<R_{\rm CR}/R_{\rm bar}<1.4$). A strong and fast-rotating bar at such early cosmic epoch is likely to be triggered by spontaneous bar instability since bars excited by galaxy interaction tend to be weak and stay slow-rotating ($R_{\rm CR}/R_{\rm bar}>1.4$) for long times\cite{1998ApJ...499..149M,2017MNRAS.464.1502M}.
\par
Our analyses show the barred morphology of J0107a is accompanied by dynamical features routinely found in observations of nearby galactic bars, namely symbolic non-circular motion indicated by the S-shape isovelocity line\cite{2005MNRAS.364..773F,2022ApJ...939...40L} (Figure 3c) and butterfly-shape torque\cite{2004AJ....128..183B,2024arXiv241013595L} (Figure 2c) funneling gas into the center\cite{2005A&A...441.1011G}. These archetypal features imply that the bar in J0107a acts in the same way as local bars but with an unprecedented inflow rate. The angular momentum loss per rotation in the order of $\sim10\%$ inside bar semi-major axis length is similar to that in local barred spirals with comparable $A_2$\cite{2009ApJ...692.1623H,2013A&A...555A.128T}, so the main driver of this inflow rate is the average molecular gas surface density of $1800\pm300$ $M_\odot$ pc$^{-2}$ in the disk, in contrast to typical value of $10-100$ $M_\odot$ pc$^{-2}$ in local disk galaxies\cite{2013AJ....146...19L}. For example, NGC1365, one of the most actively star-forming nearby barred spirals, has an $A_2$ of 0.7 \cite{2016A&A...587A.160D}, molecular gas inflow rate of 50 $M_\odot$ yr$^{-1}$ \cite{2009ApJ...703.1297E} from the bar and total SFR of 17 $M_\odot$ yr$^{-1}$ \cite{2023ApJ...944L..19L}, while J0107a has size half of NGC1365 but $\sim10\times$ more molecular gas, resulting in $\sim10\times$ higher inflow rate and an intense starburst of SFR$\approx500$ $M_\odot$ yr$^{-1}$. Therefore, the behavior of high-redshift bars can be much more violent than local ones. 
\par
The confirmation of bar-induced dynamical processes (Figure 4) in J0107a marks the earliest onset of secular evolution in the heaviest disk galaxies. The ALMA observations also detected an extended molecular gas halo and tentative filamentary structures out to $\approx30$ kpc radius (Extended Data Figure 2). While deeper observations are needed to investigate the nature of the diffuse metal-enriched cold gas surrounding J0107a, the abundant extended gas can be either a signature of accretion from the cosmic web\cite{2009Natur.457..451D} or gas recycling in the circumgalactic medium\cite{2017ARA&A..55..389T}.  Inside the stellar disk, the molecular gas fraction $M_{\rm mol}/M_\star\approx0.7$ is much higher than the typical value of $\leq0.1$ in local $M_\star\sim10^{11}M_\odot$ barred galaxies\cite{2019PASJ...71S..14S}. The bulge of J0107a has an exponential radial profile (Extended Data Figure 7), similar to disky bulges (a.k.a. pseudobulges) in local barred spirals\cite{2004ARA&A..42..603K}. Together with the intense starburst in the bar region, our results imply that a significant fraction of stellar mass will be added to the bar and/or pseudobulge in future. We also see a lower gas velocity dispersion in the core of J0107a (Figure 3e/f), which can be the progenitor of the low velocity dispersion stellar component in the centers of local barred galaxies\cite{2006MNRAS.369..529F}, formed at $z\sim2$\cite{2015A&A...584A..90G}. Therefore, J0107a is a young system and may represent the formation phase of present-day bars and pseudobulges. The fact that J0107a is the most distant and massive strongly barred galaxy known to date is in line with galaxy downsizing\cite{1996AJ....112..839C}: more massive disks have formed and settled down earlier, allowing formation of bars\cite{2008ApJ...675.1141S}.
\par
The early onset of a large bar in J0107a's baryon-dominated (baryon mass fraction $=0.8_{-0.3}^{+0.2}$ within 2.2 effective radii) disk align with the trends found by bar formation models, but the abundant gas brings challenges to them. Both cosmological and idealized simulations suggest a higher baryon fraction inside the disk promotes bar formation\cite{2020MNRAS.491.2547R,2022MNRAS.512.5339R,2023ApJ...947...80B,2024A&A...687A..53V,2024arXiv240906783F,2024ApJ...968...86B}. Some of these simulations have found star-forming bars at high redshifts\cite{2022MNRAS.512.5339R,2024ApJ...968...86B}, and our observations of a real high-redshift bar points to a link between bar-driven gas dynamics and active star formation. On the other hand, the high gas fraction of $M_{\rm mol}/M_\star\approx0.7$ in a fully developed ($A_2\sim0.8$) stellar bar is unexpected. An extrapolation from the trend in the IllustrisTNG cosmological hydrodynamic simulation predicts $M_{\rm mol}/M_\star<0.1$ (upper $80\%$ bound) for barred galaxies at this redshift and stellar mass\cite{2022MNRAS.512.5339R}.  The steep decrease of $M_{\rm mol}/M_\star$ as a function of stellar mass in simulated high-redshift bars suggests implementations of feedback as the origin of this discrepancy since feedback removes gas from the bar region\cite{2024arXiv240906783F} and alters disk dynamics\cite{2024arXiv241116876R}. At the beginning of bar formation in J0107a, the gas fraction could be even higher ($M_{\rm mol}/M_\star>4$ at $z>3$ for constant SFR and total mass), while galaxies in the simulation always have $M_{\rm mol}/M_\star<0.7$ at the time of bar formation\cite{2021A&A...647A.143L}. Similarly, controlled simulations of isolated disks have found a high gas fraction delays bar formation and weakens bars\cite{2010ApJ...719.1470V,2013MNRAS.429.1949A}. 
Very recent controlled experiments of Milky Way-mass systems have shown bar-like structures do form in gas-rich disks with radial flows as a driver\cite{2024ApJ...968...86B}, but their bars still lessen strength progressively at higher gas fraction and dissolve for initial $M_{\rm mol}/M_\star\geq1.5$. Whether their conclusion holds in a $>10\times$ deeper potential well (Figure 2b) like J0107a needs test, and the diffuse CO(4\textendash3) and [CI](1\textendash0) emission surrounding J0107a suggests cold accretion as a possible missing piece in controlled simulations. In summary, our observations show that a classical bar can grow and become mature in early gas-rich disks and highlight the need for refining bar formation models at high gas fractions.
\par
Besides the striking structural and dynamical similarities with local barred spirals, J0107a is a bright submillimeter galaxy (SMG) with an 888 $\mu$m flux density of 8 mJy. The SFR and gas fraction are typical of a massive star-forming galaxy at $z\sim2.5$ \cite{2018ApJ...853..179T}, the peak of cosmic star formation. Massive and dusty star-forming galaxies at high redshifts can be easily detected at submillimeter wavelengths, making up the SMG population\cite{2020RSOS....700556H}, and the physical mechanisms that bring a large amount of gas into their central star-forming regions are the key for understanding their role in galaxy evolution. Bars have been suspected to act as a gas transporter in SMGs\cite{2019ApJ...876..130H}, and our new VLA and ALMA observations show bar can indeed power a bright SMG. This is also consistent with the diverse formation channels of SMGs suggested by observations and theoretical studies\cite{2013MNRAS.428.2529H,2024A&A...691A.299G}: both galaxy merger/interaction and secular gas-rich disk can produce an SMG. For the latter, bars might be an important ingredient because of increasing evidence for their presence in SMGs or similarly massive, high-SFR galaxies at $z>2$\cite{2019ApJ...876..130H,2023ApJ...945L..10G}.

\newpage
\begin{figure}
    \center
    \includegraphics[width=0.9\linewidth]{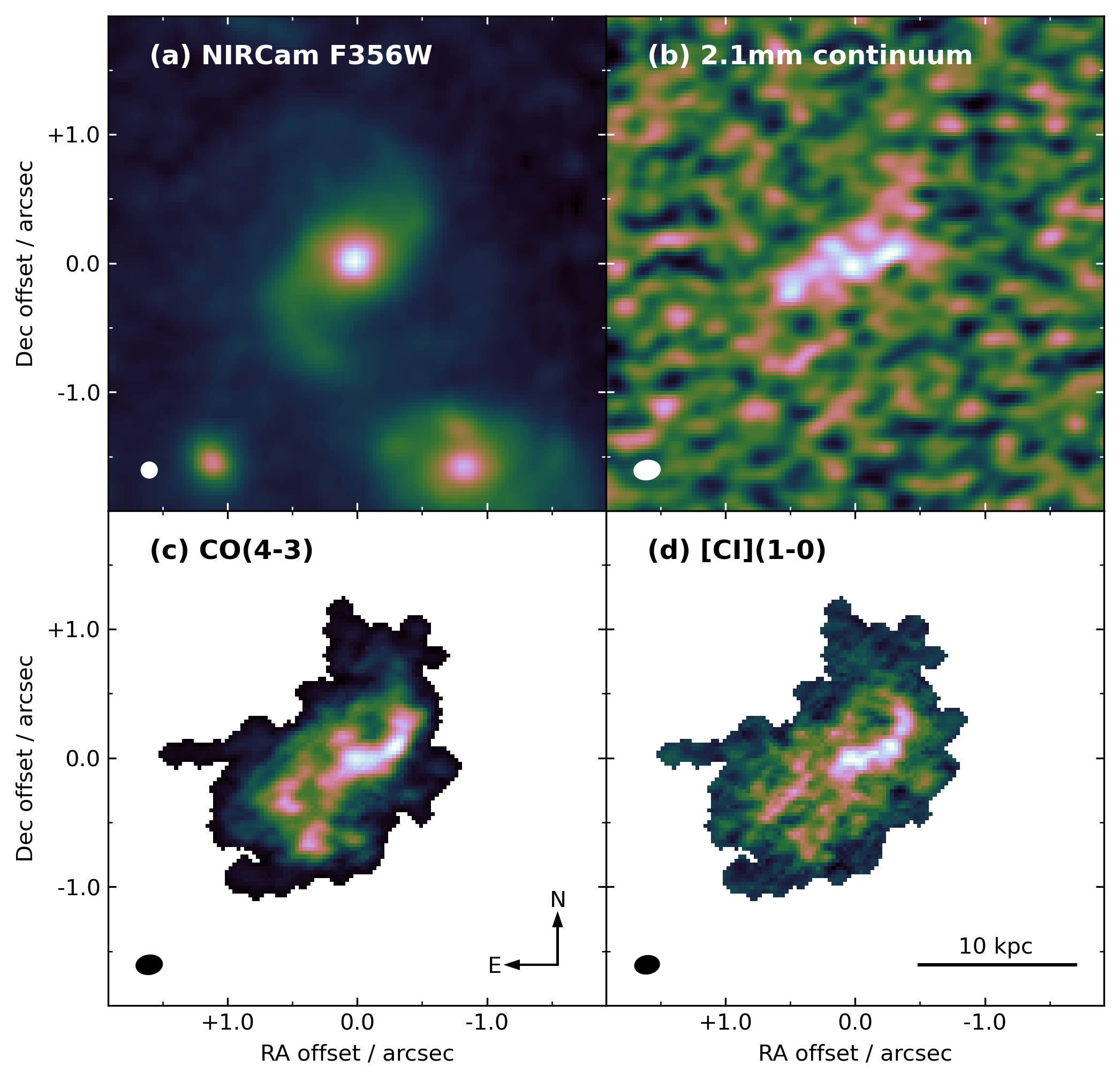}\\
    {{\bf Figure 1 $|$  The giant barred spiral J0107a at $z=2.467$} {\bf (a)} JWST NIRCam F356W image; {\bf (b)} ALMA 2.1mm continuum image; {\bf (c)} CO(4\textendash3) velocity-integrated line intensity; {\bf (d)} [CI](1\textendash0) velocity-integrated line intensity. The ellipse in the lower left of each panel shows the beam FWHM. The southwest spiral galaxy is in foreground ($z=1.186$). The elliptical galaxy southeast to J0107a is detected neither in line nor in continuum emission and likely at a different redshift (photometric redshift $z\sim1.7$). }
    \label{fig1}
\end{figure}
\newpage
\begin{figure}
    \center
    \includegraphics[width=0.9\linewidth]{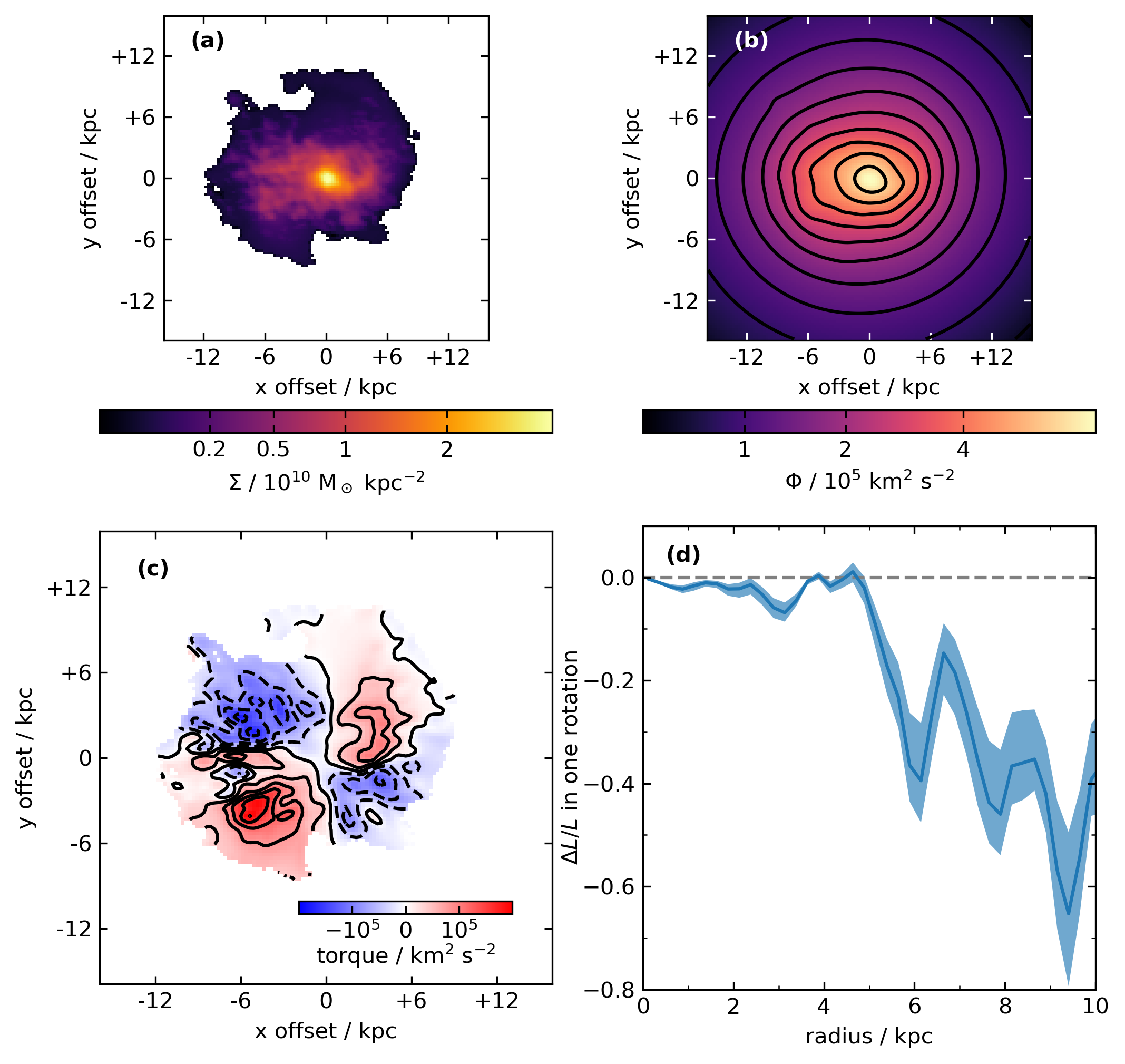}\\
    {{\bf Figure 2 $|$  Calculation of gravitational torque }{\bf (a)} Mass surface density map of J0107a de-projected for position and inclination angles; {\bf (b)} gravitational potential map; {\bf (c)} gravitational torque map; {\bf (d)} fractional angular momentum change of the molecular gas per rotation as a function of radius. The shaded region indicates $1\sigma$ uncertainty. Negative values of $\Delta L/L$ indicate loss of angular momentum and hence inward flow. Detailed steps of the calculation are described in method.}
    \label{fig2}
\end{figure}
\newpage
\begin{figure}
    \center
    \includegraphics[width=0.99\linewidth]{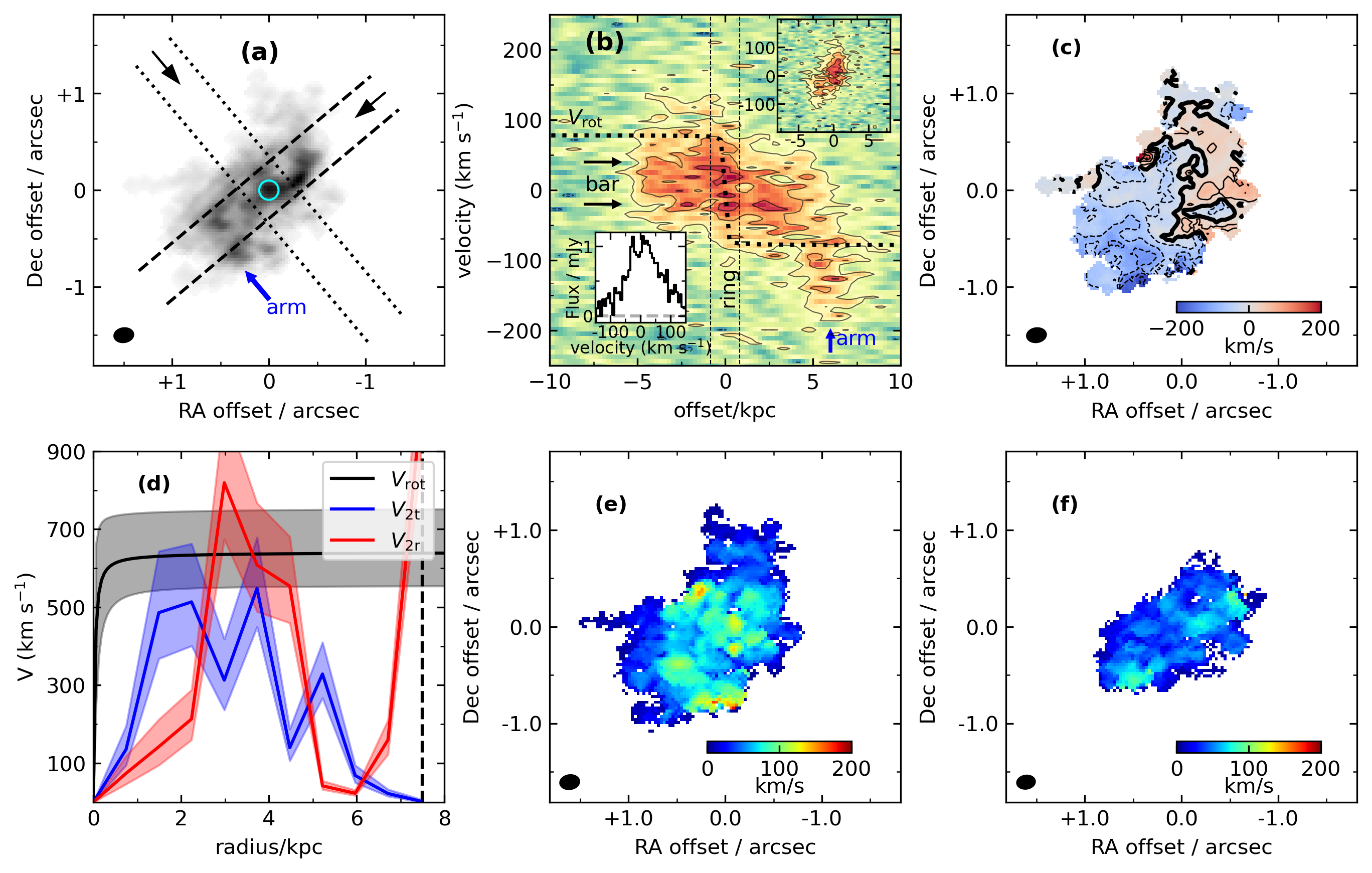}\\
    {\bf Figure 3 $|$  The velocity field of J0107a. }{\bf (a):} Slits used to extract PV diagram from CO(4\textendash3) data along the major axis (dashed lines and arrow) and minor axis (dotted lines and arrow) overplotted on moment 0 map; {\bf (b):} PV diagram along the major axis, contour starts from $\pm2\sigma$ with $2\sigma$ steps. The insert panels show the spectrum extracted from a $0\farcs2$ diameter aperture placed on the central peak (the cyan circle in {\bf(a)}) in lower left and PV diagram along the minor axis in upper right; {\bf (c):} CO(4\textendash3) LOS velocity map. Contours starts from 0 km s$^{-1}$ (thick line) with steps of $\pm20$ km s$^{-1}$;  {\bf (d):} circular ($V_{\rm rot}$) and non-circular ($V_{2\rm t}$ and $V_{2\rm r}$) velocity as a function of radius. The solid line and shaded region indicate the median and $1\sigma$ error. The median of fitted circular velocity is shown as the dotted line in panel {\bf (b)}; {\bf (e)} velocity dispersion map of CO(4\textendash3); {\bf (f)} velocity dispersion map of [CI](1\textendash0). Note that the two velocity dispersion maps are made from different masks so the absolute values cannot be compared directly. 
    \label{fig3}
\end{figure}
\newpage
\begin{figure}
    \center
    \includegraphics[width=0.9\linewidth]{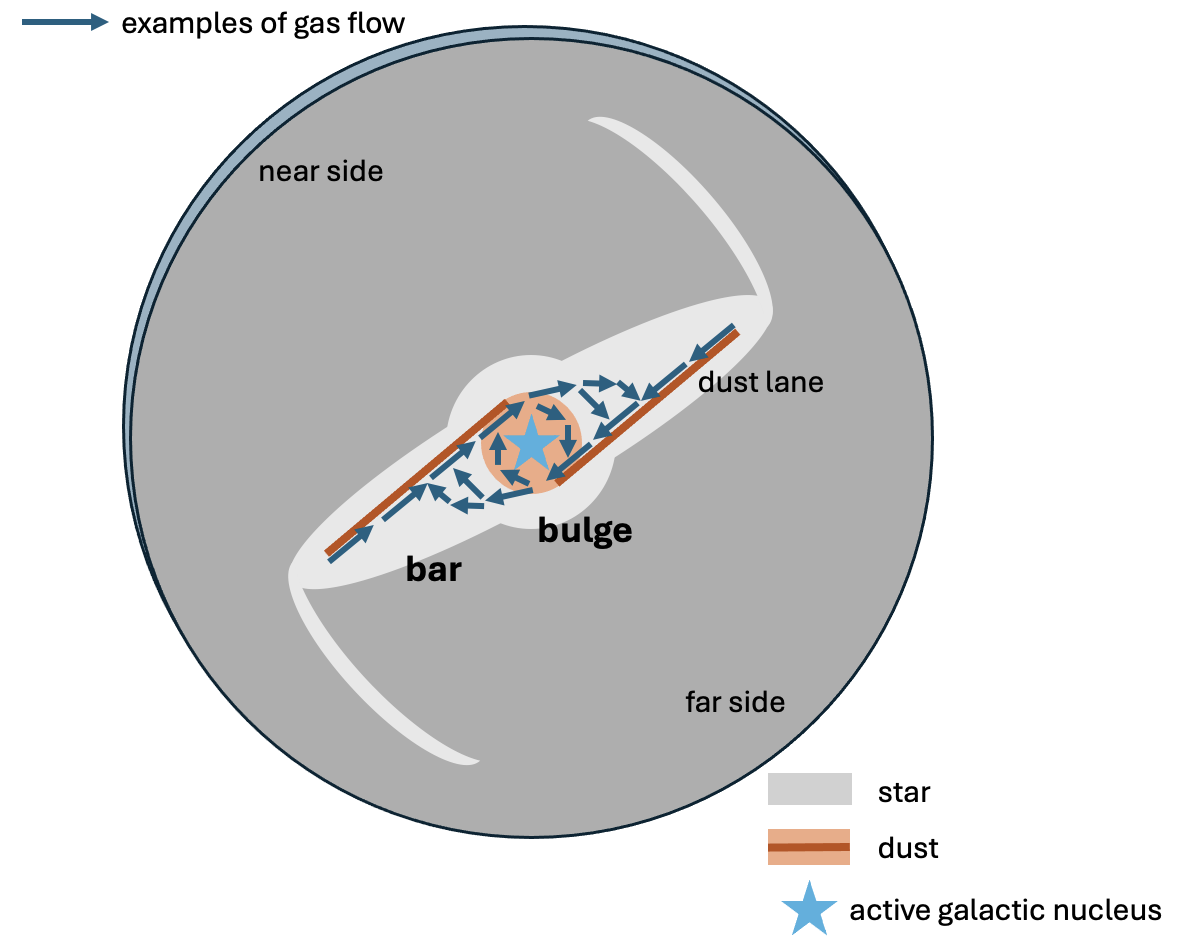}\\
    {{\bf Figure 4 $|$  Overview of J0107a. } {Schematic diagram of J0107a. The gas flows in the bar potential are denoted by blue arrows.}}
    \label{fig4}
\end{figure}







\newpage
\begin{addendum}
 \item
 We thank Akihiko Hirota and Fumiya Maeda for discussions on local bars. We are grateful to Shun Ishii for his help in ALMA observations. This work is partly supported by NAOJ ALMA Scientific Research Grant Code 2024-26A and JSPS KAKENHI Grant Numbers JP22H04939, JP23K20035, and JP24H00004. This paper makes use of the following ALMA data: ADS/JAO.ALMA \#2023.1.01262.S. ALMA is a partnership of ESO (representing its member states), NSF (USA), and NINS (Japan), together with NRC (Canada), MOST and ASIAA (Taiwan), and KASI (Republic of Korea), in cooperation with the Republic of Chile. The Joint ALMA Observatory is operated by ESO, AUI/NRAO and NAOJ. This paper uses data from the Karl G. Jansky Very Large Array operated by the National Radio Astronomy Observatory under project VLA/23B-024. The National Radio Astronomy Observatory is a facility of the National Science Foundation operated under cooperative agreement by Associated Universities, Inc.
\item[Author Contributions]
SH proposed the VLA and ALMA observations, analyzed the data, and wrote the manuscript. RK, HU, KK, YT, and TS are responsible for the discovery of the target and contribute to the interpretation and discussion of the data.
\item[Competing interests] The authors declare no competing interests.
\item[Data availability] The data analyzed in this paper are publicly available in JWST, VLA and ALMA archives.
\item[Code availability] We used standard data analysis tools: astropy, photutils, numpy, scipy and 3DBarolo, all of which are publicly available on the internet.
\item[Correponding Author] Shuo Huang (shuohuang.version3@gmail.com)
\end{addendum}


\newpage 
\begin{methods}
\subsection{Cosmology and Initial mass function}
We adopt Planck 2018 cosmological parameters: $H_0=67.4$ km s$^{-1}$ Mpc$^{-1}$, $\Omega_{m}=0.315$ and $\Omega_{\mathrm{\Lambda}}=0.685$\cite{2020A&A...641A...1P} and  the Chabrier (2003)\cite{2003PASP..115..763C} initial stellar mass function. Errors of scalar values represent $1\sigma$ confidence interval (16th, 50th and 84th percentiles).
\end{methods}

\subsection{ALMA observation and data processing}
ALMA observations of the CO(4\textendash3) and [CI](1\textendash0) emission lines were conducted simultaneously in one spectral tuning from November to December 2023 (project code 2023.1.01262.S, principle investigator: Shuo Huang) using the Band 4 receiver\cite{2014PASJ...66...57A} in C-7 hybrid and C-4 array configurations. The on-source integration time was 3.38 hours and 1.16 hours, respectively. The data are reduced using the standard ALMA pipeline in the Common Astronomy Software Applications (CASA) package. Spectral cubes were generated using the \texttt{tclean} task in CASA with Briggs weighting (robust=0.5) and 3-channel binning, resulting in a beam size of $0\farcs19\times0\farcs14$ with a position angle (PA) of $82\arcdeg$ and an RMS noise of 0.2 mJy beam$^{-1}$ at velocity resolution of 6.6 km s$^{-1}$ before primary beam correction for CO(4\textendash3). The continuum is estimated and subtracted using the \texttt{imcontsub} task in CASA. The channel map of CO(4\textendash3) is shown in Extended Data Figure 1. To enhance the signal-to-noise ratio (S/N) of extended structures, we also make cubes with $0\farcs3$ \texttt{uvtaper} and natural weighting, resulting in beam size of $0\farcs8\times0\farcs6$, close to the native resolution of the C-4 array.
\par
For the 2.1 mm continuum, we add data from partially failed observations conducted with C-8 array that resulted in no usable spectral line data. The combined data has a total on-source integration time of 7.93 hours and covers the sky frequency range of 142.42\textendash144.28 GHz, only 1/4 bandwidth of default continuum applications. Continuum image is generated using the \texttt{tclean} task with multi-frequency synthesis algorithm and Briggs weighting (robust=0.5). The beam size is adjusted to match the CO(4\textendash3) cube by specifying the \texttt{restoringbeam} parameter. The final continuum image has an RMS noise level of 6.3 $\mu$Jy beam$^{-1}$.
\begin{figure}
    \center
    \includegraphics[width=0.9\linewidth]{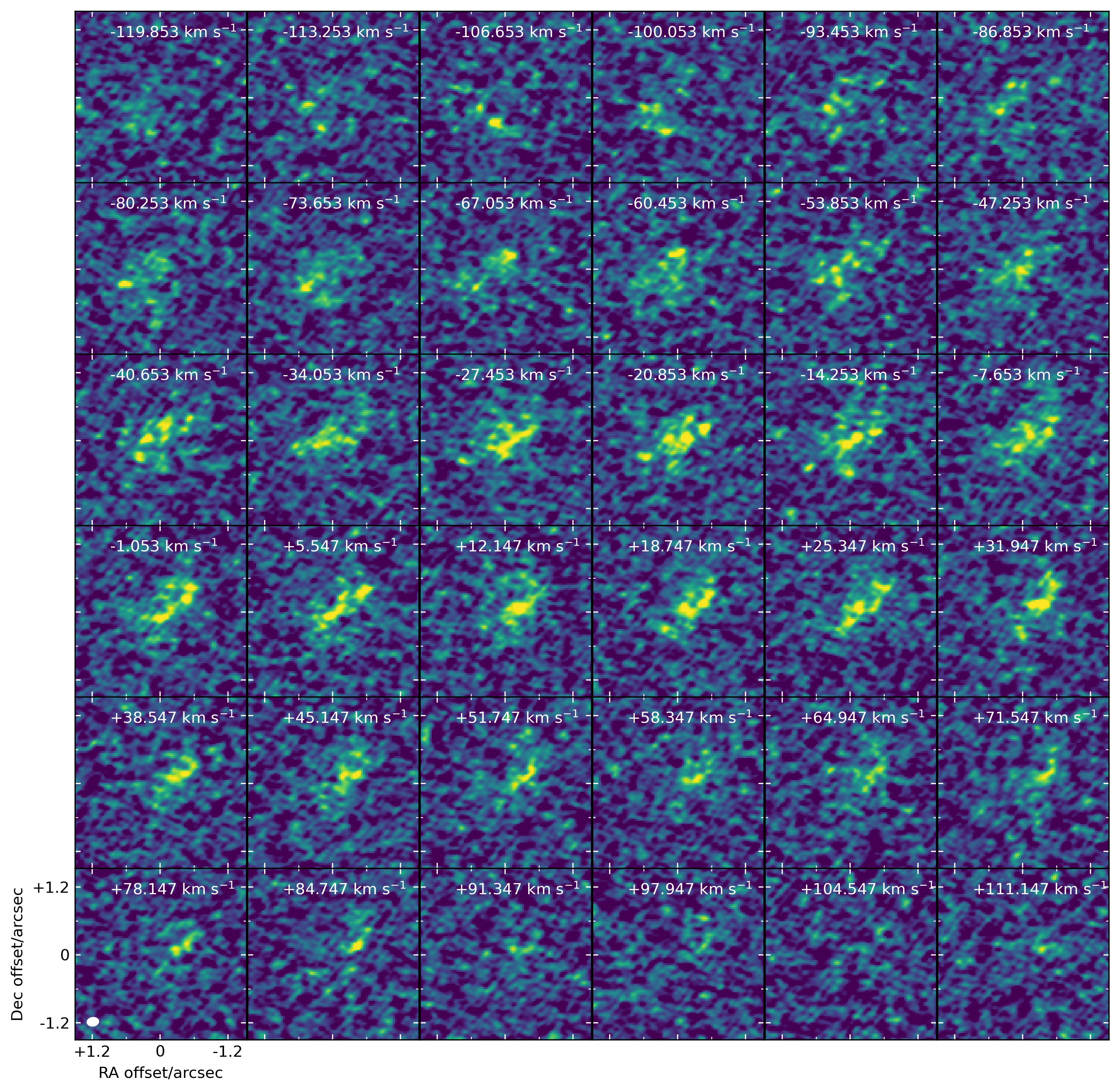}\\
    {{\bf Extended Data Figure 1 $|$ }  The channel map of the CO(4\textendash3) emission line. The global velocity gradient at PA$\approx310\arcdeg$ and non-circular motion are visible.}
    \label{fig:s1}
\end{figure}
\par
The CO(4\textendash3) and [CI](1-0) spectral cubes are processed following the method of PHANGS-ALMA pipeline\cite{2021ApJS..255...19L}. Signal masks are generated by finding continuously connected pixels with an area larger than half a beam and S/N$>4$ in at least two consecutive channels. These core masks are extended to surrounding pixels with S/N$>2$ in more than two consecutive channels and further dilated by 2 pixels in spatial and 2 channels in spectral direction to include all low S/N wings. Integrated line intensity (the zeroth moment), intensity-weighted line-of-sight velocity (the first moment), velocity dispersion (the second moment) and the corresponding error maps are then computed after applying the masks to the cube. For the fainter [CI](1-0) line, we combine the mask with that from CO(4\textendash3) to generate moment 0 and 1 maps. The moment 2 map of [CI](1\textendash0) is made from [CI]-only mask to reduce noise. In Extended Data Figure 2, we show the moment 0 maps of \texttt{uvtaper}ed cubes. No companion galaxy is detected in the cube or continuum image (beam half power full width = 40 arcsec, equivalent to a diameter of $\approx330$ physical kpc at $z=2.467$).
\begin{figure}
    \center
    \includegraphics[width=0.9\linewidth]{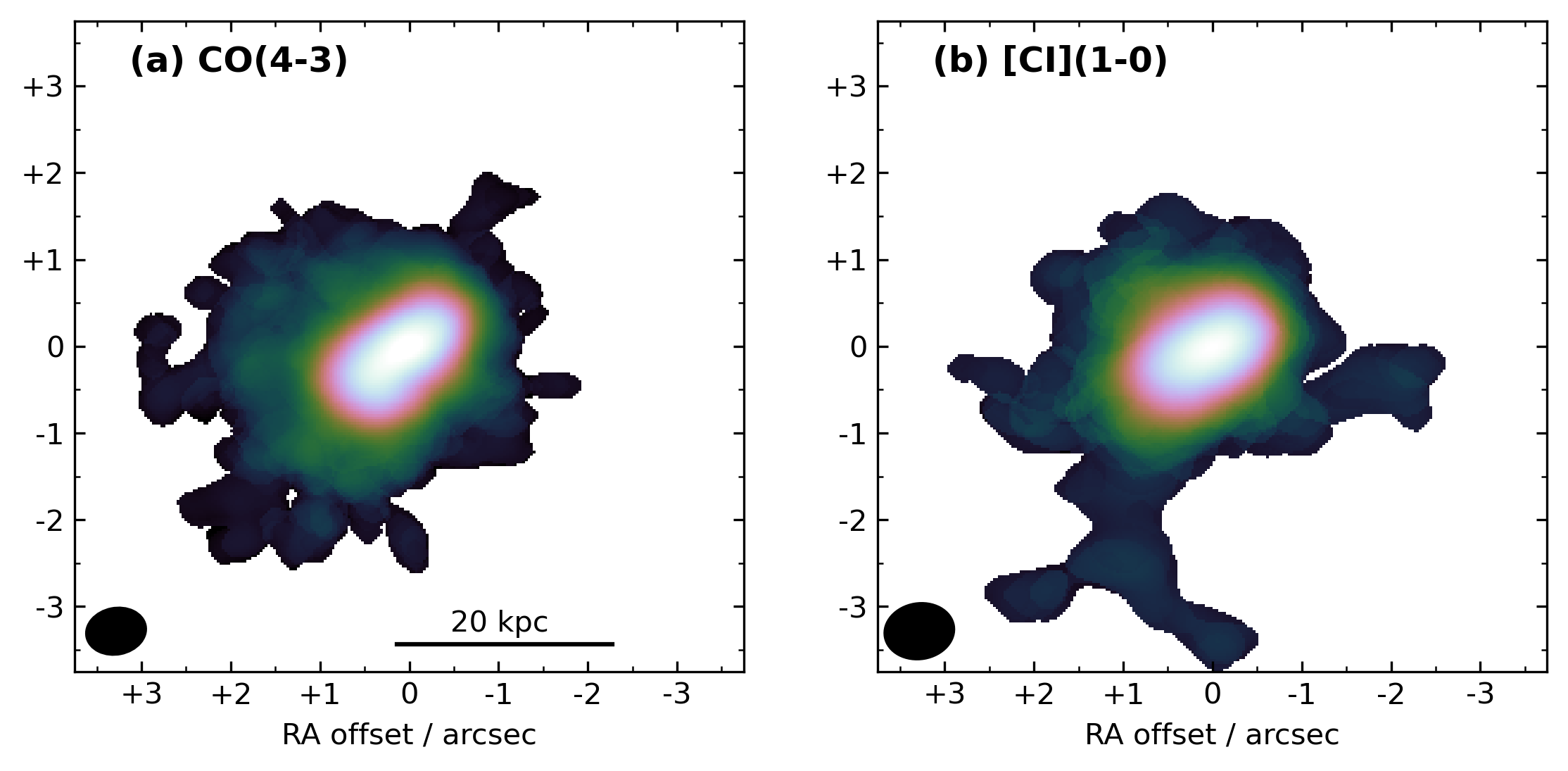}\\
    {{\bf Extended Data Figure 2 $|$ }{\bf (a):} low-resolution CO(4\textendash3) line intensity map to highlight the extended emission;
     {\bf (b)}: same as {\bf (a)} but for [CI](1\textendash0).}
    \label{fig:s2}
\end{figure}
\par
The center of the galaxy is determined using the high-resolution cube and zeroth moment map of the CO(4\textendash3) emission. By fitting a 2-dimensional Gaussian to the central peak, we derive the sky coordinate of as right ascension 01h07m48.30s and declination -17d30m28.027s in the ICRS frame. The system redshift is determined by Gaussian fitting to the 1-dimensional CO(4\textendash3) spectrum of the central peak, yielding $z=2.4669$. These parameters are fixed during subsequent analysis.
\subsection{VLA observation}
We conducted VLA observation of J0107a to detect the CO(1\textendash0) emission line on November 13, 2023 (project VLA/23B-024, principle investigator: Shuo Huang). We use the Ka-band receiver and 8-bit sampler in the D-array configuration. The Wideband Interferometric Digital ARchitecture (WIDAR) correlator is configured to place two basebands centered at the sky frequency of 33.19 GHz and 34.88 GHz to cover the redshifted CO(1\textendash0) emission line from J0107a. The on-source integration time was 1.68 hours. The data were reduced using the standard VLA pipeline in the CASA package. Spectral cube was generated using the \texttt{tclean} task in CASA with Briggs weighting (robust=0.5) and 18.0 km s$^{-1}$ channel width. The resulting spectral cube has a beam size of $2\farcs9\times1\farcs8$ with a PA of $1\arcdeg$ and an RMS noise of 0.2 mJy beam$^{-1}$ before primary beam correction. The 34 GHz radio continuum of J0107a remains undetected down to $1\sigma=87$ $\mu$Jy beam$^{-1}$.

\begin{figure}
    \center
    \includegraphics[width=0.8\linewidth]{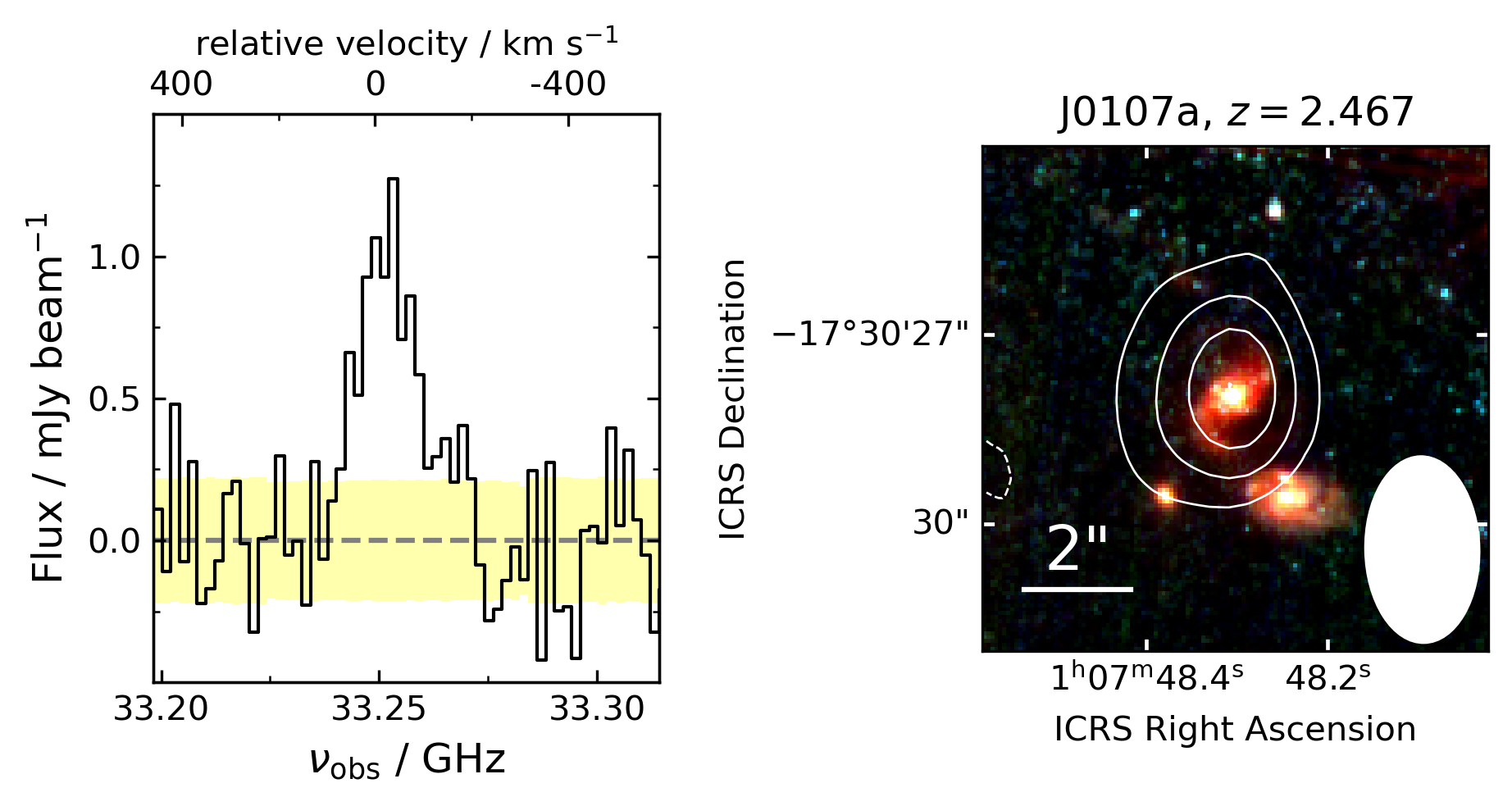}\\
    \includegraphics[width=0.8\linewidth]{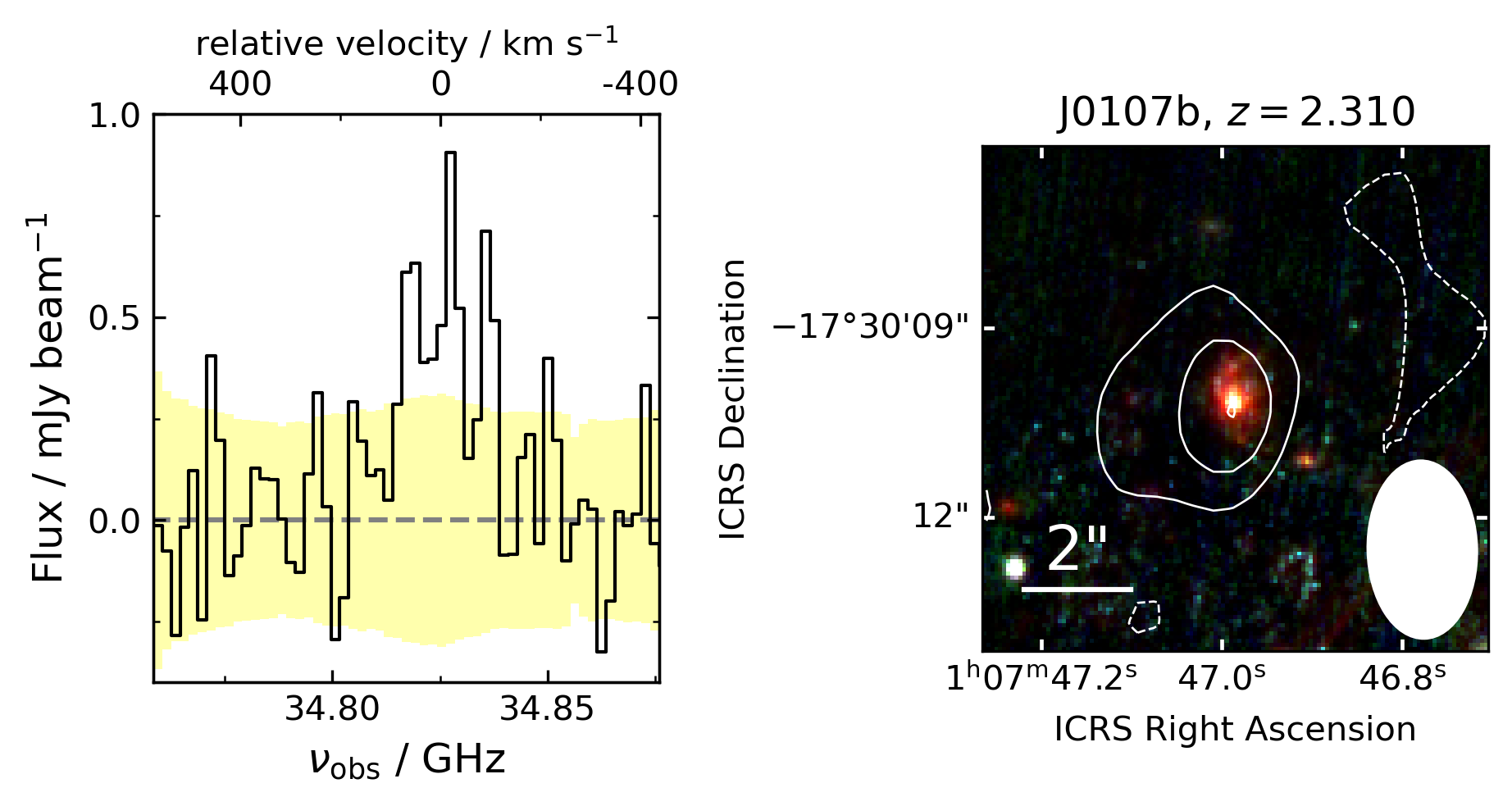}\\
    {{\bf Extended Data Figure 3 $|$} {\bf Top:} The CO(1\textendash0) spectrum (left) and JWST NIRCam image (RGB = F356W, F200W, F150W)} with CO(1\textendash0) velocity-integrated intensity overlaid as white contours (right) of J0107a. The light yellow region indicates the RMS noise level. {\bf bottom:} the same as top but for J0107b. Contours start from $\pm2\sigma$ with $2\sigma$ steps. 
    \label{fig:s3}
\end{figure}
\subsection{Molecular gas mass}
The CO(1\textendash0) emission line from J0107a is detected at a peak S/N of 6 in the channel maps (Extended Data Figure 3, top panels). From the zeroth moment map made by collapsing spectral channels containing the emission line, we measure a velocity-integrated CO(1\textendash0) flux density of $0.27\pm0.04$ Jy km s$^{-1}$, which translates to line luminosity $L^{'}_{\rm CO(1-0)}=(8.1\pm1.1)\times10^{10}L_\odot$. Since J0107a is a seemingly isolated regular rotating disk with a modest SFR for its huge mass, we use conversion factor $\alpha_{\rm CO}=M_{\rm mol}/L^{'}_{\rm CO(1-0)}=3.6_{-0.5}^{+0.6}$ $M_\odot$ (K km s$^{-1}$ pc$^{2}$)$^{-1}$ from the scaling relations between stellar mass, metallicity and $\alpha_{\rm CO}$\cite{2018ApJ...853..179T} for general high-redshift star-forming galaxies, where the error includes CO(1\textendash0) flux and stellar mass uncertainties and the scatter of scaling relations. The $M_{\rm mol}$ from CO(1\textendash0) is $M_{\rm mol} =(3.3\pm0.6)\times10^{11}M_\odot$ and consistent with $M_{\rm mol}=(3.2\pm1.6)\times10^{11}M_\odot$ based on dust continuum\cite{2021ApJ...917...94M} and $M_{\rm mol}=(3.3\pm0.6)\times10^{11}M_\odot$ from [CI](1-0) and recent cross-calibrated conversion factor\cite{2022MNRAS.517..962D}.
\par
J0107b, another SMG at $z=2.310$ with an 888 $\mu$m flux of 5 mJy, 26$\arcsec$ northwest from J0107a\cite{2021ApJ...917...94M}, is also detected by our VLA observation with CO(1\textendash0) velocity-integrated flux density of $0.20\pm0.05$ Jy km s$^{-1}$ (Extended Data Figure 2, bottom panels). Similarly, J0107b has $M_{\rm mol} =(2.4\pm0.7)\times10^{11}M_\odot$. Following the same procedures as J0107a\cite{2023ApJ...958L..26H}, we derive $M_\star = 1.6^{+0.5}_{-0.5}\times10^{11}M_\odot$ and SFR$=144^{+93}_{-55}M_\odot$ yr$^{-1}$ for J0107b. J0107b has a normal spiral shape in NIRCam images. Both J0107a and J0107b follow the main sequence scaling relations between $M_\star$, SFR and $M_{\rm mol}$\cite{2018ApJ...853..179T}, but they are not physically associated given the redshift difference of $\approx0.15$.

\subsection{Spatial configuration of J0107a's disk}
Based on the spiral arm shape and moment 1 map, we can determine that the galaxy rotates clockwise, and north is the near side.
The inclination $i$ and major axis PA are the basis of kinematic analysis, but the strong bar distortion prevents these two key quantities from being reliably derived from high-resolution data. The photometric methods rely on ellipse fit to the light profiles. This cannot be done on the JWST images because of strong distortion by the spiral arms and lacking sensitivity to the underlying smooth disk out to large radii. Nonetheless, the outermost isophote of J0107a has an ellipticity of 0.147 and sets an upper limit of $i\leq30\arcdeg$. Alternatively, one can estimate $i$ and PA by fitting a rotating disk model to the spectral cube. To minimize the influence of the bar, we make a $0\farcs7$ uvtapered cube and Briggs r=0.5 weighting, then fit it using \texttt{3DBarolo} code\cite{2015MNRAS.451.3021D}. Since the S-shape twist is still visible (Extended Data Figure 4a), we let the code model radial inflow using the \texttt{VRAD} parameter. The \texttt{SPACEPAR} task is used to explore the parameter space between $1\arcdeg<i<60\arcdeg$ and $250\arcdeg<$PA$<350\arcdeg$, given previous estimates of $i\approx10\arcdeg$ and PA$\approx295\arcdeg$\cite{2023ApJ...958L..26H}. The results (Extended Data Figure 3b/c/d) show that the inclination converges to $i\approx7\arcdeg$, while the PA is poorly constrained with best-fit ranges in $290\arcdeg-320\arcdeg$ and depends on the initial guess on \texttt{VRAD}. Both values have substantial uncertainties, and we will return to the inclination later. For the loose range of the PA, we attribute this to the presence of an S-shape twist (i.e., a non-zero \texttt{VRAD}) making it difficult to determine the true orientation of the minor axis, along which pure circular motion velocity is zero. In the first moment map of the \texttt{uvtaper}ed cube, the zero-velocity line crossing the center gives a wrong PA of $300\arcdeg$, which is close to the best-fit value from \texttt{3DBarolo}. Instead, assuming the bar influence becomes small at large radii (see also Extended Data Figure 6), a line crossing the ends of the S-shape twist gives PA$=310\arcdeg$, which also agrees with the global velocity gradient in the channel map (Extended Data Figure 1) but not perpendicular to the zero velocity line in high-resolution data (Figure 3e). We adopt this value as the best-effort estimate of the PA.
\begin{figure}
   \center
    \includegraphics[width=0.9\linewidth]{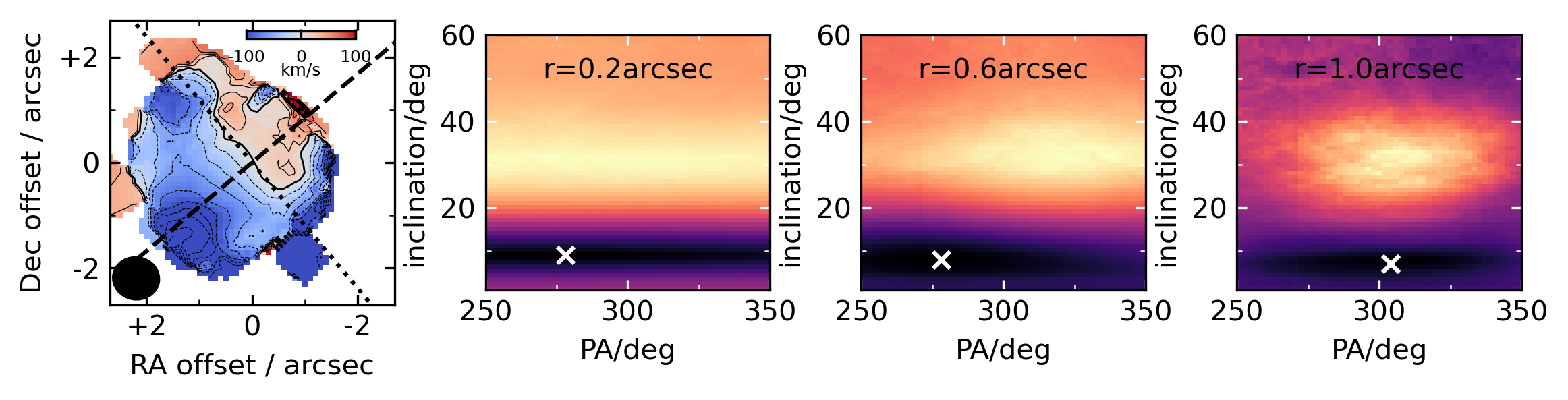}\\
    {{\bf Extended Data Figure 4 $|$ }{\bf (a)} $0\farcs7$ \texttt{uvtaper}ed CO(4\textendash3) LOS velocity map. The orientation of major and minor axes are marked as dashed and dotted lines, respectively. {\bf (b)\textendash(d)}: results of parameter space exploration using \texttt{3DBarolo}. Darker means higher likelihood. The best-fit value is marked as the white cross.}
    \label{fig:s4}
\end{figure}

\subsection{2D velocity field fitting}
To estimate the degree of non-circular motion, we fit the observed CO(4\textendash3) velocity field using a combination of circular and $m=2$ Fourier mode models\cite{2007ApJ...664..204S}. The circular part is:
\begin{equation}
 V_{\rm circ}(r) = V_{\rm rot}(r)\cos\theta_0\sin{i},
\end{equation}
where $\theta_0$ is the angle measured counter-clockwise from the disk PA, and $i$ is the inclination angle. The functional form of radius $r$ for rotation velocity is parameterized as $V_{\rm rot}=2V_{\rm max}\arctan(r/r_t)/\pi$ with two free parameters $V_{\rm max}$ and $r_t$ as frequently done in extragalactic studies.
The non-circular part is modeled as \begin{equation}
 V_{\rm noncirc}(r) = -V_{\rm 2t}(r)\cos2\theta_1\sin{i}- V_{\rm 2r}(r)\sin2\theta_1\sin{i},
\end{equation}
where $V_{2t}$ and $V_{2r}$ are the azimuthal and radial components of the non-circular motion and are allowed to vary freely as a function of radius. $\theta_1$ is the angle measured from the PA of the non-circular motion and may not be identical to the PA of the bar\cite{2024ApJ...973..116D}.
\par
For circular motion, the velocity $V_{\rm rot}$ is highly dependent on inclination. An inclination of 10$\arcdeg$ causes little difference in the visual appearance of the galaxy (axis ratio changes by $<1\%$), but the dynamical mass decreases by half.
This implies that dynamical arguments cannot be applied to constrain light-to-mass ratios since it is sensitive to even a subtle change in $i$ and accurate measurement of a low $i$ value is difficult. While substantial likelihood exists for inclination values $i>10\arcdeg$ in the \texttt{3DBarolo} fit (Extended Data Figure 3), their dynamical masses can not be reconciled with $M_\star$ and $M_{\rm mol}$ within error even for $\alpha_{\rm CO}=0.8$ $M_\odot$ (K km s$^{-1}$ pc$^{2}$)$^{-1}$. We set the uncertainty range of $i$ to be $\pm2\arcdeg$, which is a conservative range because $i=5\arcdeg$ translates to $V_{\rm circ}\approx1000$ km s$^{-1}$ and that a single $z\sim2.5$ galaxy inhabits a dark matter halo of mass $\approx10^{14.1\pm0.7}M_\odot$\cite{2019MNRAS.483L..98K}. This is problematic because (1) it is above the $3\sigma$ bound of all-sky maximum halo mass predicted by $\Lambda$CDM\cite{2023MNRAS.518.2511L} and has theoretical prediction at $z=0$ of $10^{16.2\pm1.2}M_\odot$\cite{2013ApJ...770...57B}, higher than the most massive galaxy clusters in the local universe\cite{2019A&A...628A..86B}; (2) it has exceeded the critical halo mass for cold accretion stream to feed galaxy growth\cite{2006MNRAS.368....2D,2022ApJ...926L..21D} by $z=3.1$ or $\approx600$ Myr before observation, and this elapsed time is comparable to the current value $M_{\rm mol}$/SFR so we would not have seen such a gas-rich and highly star-forming SMG. For the upper bound $i=9\arcdeg$, it necessitates $\alpha_{\rm CO}=0.8$ $M_\odot$ (K km s$^{-1}$ pc$^{2}$)$^{-1}$ for the whole galaxy that is in tension with other molecular gas tracers, and there is no evidence for a local merger-like $\alpha_{\rm CO}$ since the gas has extended distribution and motion governed by gravity. Also, as we will discuss in the following text, the resulting molecular gas mass indicates the gas disk will collapse within one dynamical timescale due to inflow, in contradiction to the observed gaseous bar structure. As such we argue that the true value of inclination should be close to $7\arcdeg$, and $5\arcdeg<i<9\arcdeg$ is a strict bound. Nevertheless, to propagate the error of $i$ into the calculation of the gas inflow rate, we assume a Gaussian distribution centered on $7\arcdeg$ with a standard deviation of $1\arcdeg$, truncating at $5\arcdeg$ and $9\arcdeg$.
The conclusions that J0107a has a fast bar and bar-induced non-circular motion dominates over rotation are not affected by the uncertainty of $i$ because all velocities scale equally according to $1/\sin i$. Under such assumption, we derive the rotation speed at $r=10$ kpc ($\approx$2.2 times effective radius) of $V_{\rm rot}=640\pm100$ km s$^{-1}$. The total enclosed mass within 10 kpc radius is $M_{\rm dyn}=r V_{\rm rot}^{2}/G=(9\pm3)\times10^{11}M_\odot$, implying a baryon mass fraction of $f_{\rm baryon}=0.8_{-0.3}^{+0.2}$. Therefore, the mass budget in the disk region is dominated by baryon. This is in agreement with numerical simulations that show a high baryon fraction promotes early bar formation\cite{2023ApJ...947...80B,2024A&A...687A..53V}.

\subsection{Molecular gas inflow due to the bar}

We discuss the molecular gas inflow by considering the gravitational torque exerted by the bar\cite{2005A&A...441.1011G}. The method is outlined below. The stellar and gas surface density maps are made by scaling the total masses with de-projected NIRCam F444W (rest-frame 1.3 $\mu$m) and [CI](1\textendash0) intensity distributions, respectively (Extended Data Figure 5). The 2-dimensional gravitational potential field is described by
\begin{equation}
 {\rm\bf\Phi}(x,y)=-G\int{\rm \Sigma}(x^\prime, y^\prime)g(x^\prime-x, y^\prime-y)dx^\prime dy^\prime
 \end{equation}
 \begin{equation}
g(x,y)=\int_{-\infty}^{+\infty}\frac{{\rm sech}^{2}(z/h)}{\sqrt{x^2+y^2+z^2}}dz
 \end{equation}
where $G$ is the gravitational constant and $\Sigma$ is the mass surface density. The scale height $h$ is $\approx 1/12\times$ the disk scale length of $\approx2.6$ kpc.
\begin{figure}
    \center
    \includegraphics[width=0.9\linewidth]{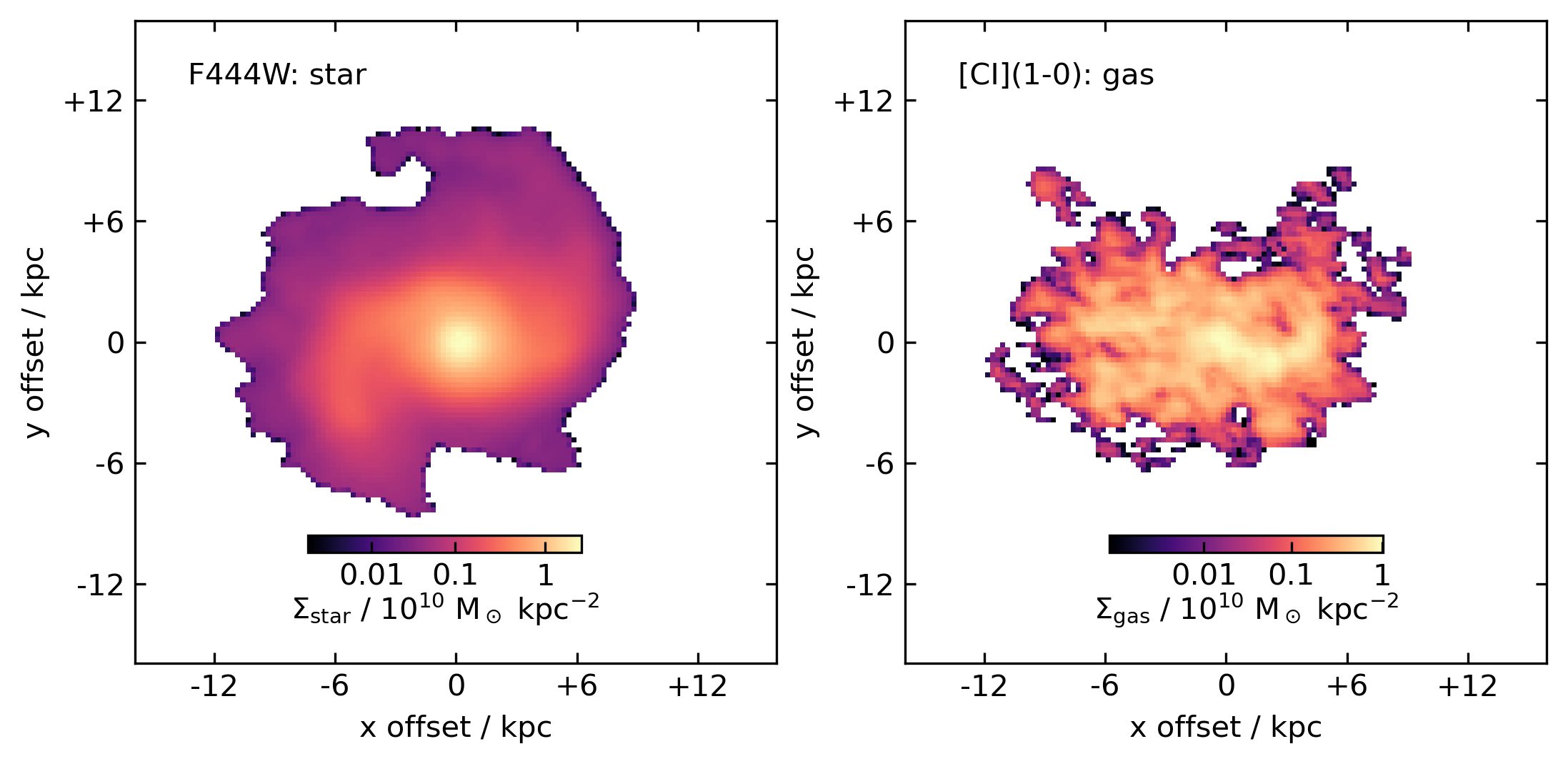}\\
    {{Extended Data Figure 5 $|$ }{\bf Left:} De-projected F444W image. {\bf right}: de-projected [CI](1\textendash0) image. Each image is shown with a logarithmic stretch.}
    \label{fig:s5}
\end{figure}
\par
The torque per unit mass at Cartesian coordinates $(x,y)$ is
\begin{equation}
 \tau(x,y)=xF_y-yF_x,
\end{equation}
where ${\bf F}=-\nabla{\rm\bf\Phi}$ is the gravitational force per unit mass. The azimuthal mean of $\tau(x,y)$ weighted by gas surface density at radius $r$, denoted as $\tau(r)$, gives a measure of instantaneous angular momentum change. The angular momentum change per rotation is
\begin{equation}
     \frac{\Delta L(r)}{L(r)} =\frac{\tau(r)T(r)}{L(r)},
\end{equation}
where $L(r)=rV_{\rm rot}(r)$ is the angular momentum of circular motion per unit mass and $T(r)=2\pi r/V_{\rm rot}(r)$ is the time for one rotation. In Figure 2d we show $\Delta L(r)/L(r)$ as a function of radius. The net gas inflow rate due to loss of angular momentum within radius $R$ is:
\begin{equation}
 \frac{dM}{dt} = \int_{0}^{R}\frac{1}{T(r)}\frac{\Delta L(r)}{L(r)}\Sigma_{\rm gas}(r) 2\pi rdr,
\end{equation}
To calculate the net inflow rate, we integrate equation (7) over the whole radial range ($R=13$ kpc) within which [CI](1-0) is detected. Because mass inflow rate scales proportionally to $\Sigma_{\rm gas}/V_{\rm rot}$, its uncertainty originates from conversions from observed flux to mass and LOS velocity to $V_{\rm rot}$, with the latter scales as $1/\sin i$. Error propagation is performed by randomly drawing $10^4$ sets of inclination, total stellar and gas masses from corresponding probability distributions and repeating the calculation of equations (3)\textendash(7). If $M_\star+M_{\rm mol}>M_{\rm dyn}$ for a given set of parameters, we cut the value of $M_{\rm mol}$ to make $M_\star+M_{\rm mol}=M_{\rm dyn}$. The resulting inflow rate is $579_{-174}^{+188}M_\odot$ yr$^{-1}$.
\par
To demonstrate the effect of the bar, the harmonic decomposition of the gravitational potential allows us to compare the radial distribution of each Fourier mode:
\begin{equation}
 {\rm\bf\Phi}(r,\theta)={\rm\bf\Phi}_0(r)+\sum_m{\rm\bf\Phi}_m(r)\cos(m\theta-\phi_m(r)).
\end{equation}
The strength of the $m$th Fourier mode is\cite{1981A&A....96..164C}:
\begin{equation}
 Q_m=\frac{m{\rm\bf\Phi}_m(r)}{r\left|\frac{d{\rm\bf\Phi}_0(r)}{dr}\right|},
\end{equation}
and the strength of the bar perturbation is measured by the ratio between maximum tangential and radial forces:
\begin{equation}
 Q_t=\frac{{\rm max}\left(\frac{\partial{\rm\bf\Phi(r,\theta)}}{\partial\theta}\right)}{r\left|\frac{d{\rm\bf\Phi}_0(r)}{dr}\right|}.
\end{equation}
The results are shown in Extended Data Figure 6. The value of $Q_t$ is above $0.4$ in $r=2-8$ kpc range, and the radial range of high $Q_2$ overlaps with strong bi-symmetric non-circular motion shown in Figure 3f. Applying Fourier decomposition directly on the stellar mass distribution, we obtain bar strength, defined as the ratio between $m=2$ and $m=0$ modes, $A_2=$0.7\textendash0.8 within the radial range $r=5.0-7.5$ kpc, which is as high as local strongly barred galaxies\cite{2016A&A...587A.160D}. Furthermore, the radial profile of the inner 4 kpc (Extended Data Figure 7) is consistent with an exponential disk, suggesting a pure pseudobulge at current spatial resolution.
\begin{figure}
    \center
    \includegraphics[width=0.5\linewidth]{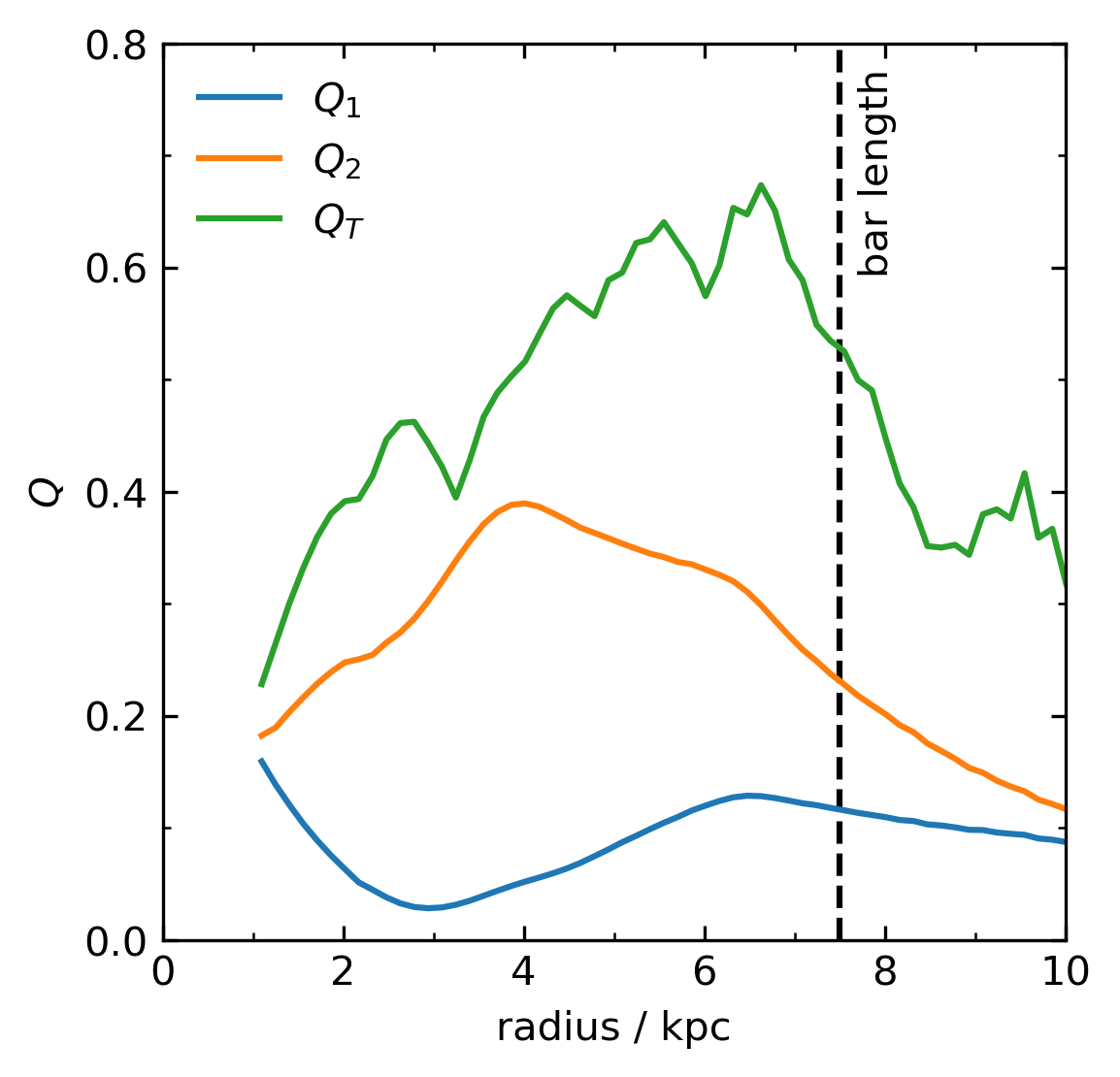}\\
    {{\bf Extended Data Figure 6 $|$ } Fourier decomposition of the gravitational potential field. The vertical dashed line marks the bar semi-major axis length.}
    \label{fig:s6}
\end{figure}
\begin{figure}
    \center
    \includegraphics[width=0.5\linewidth]{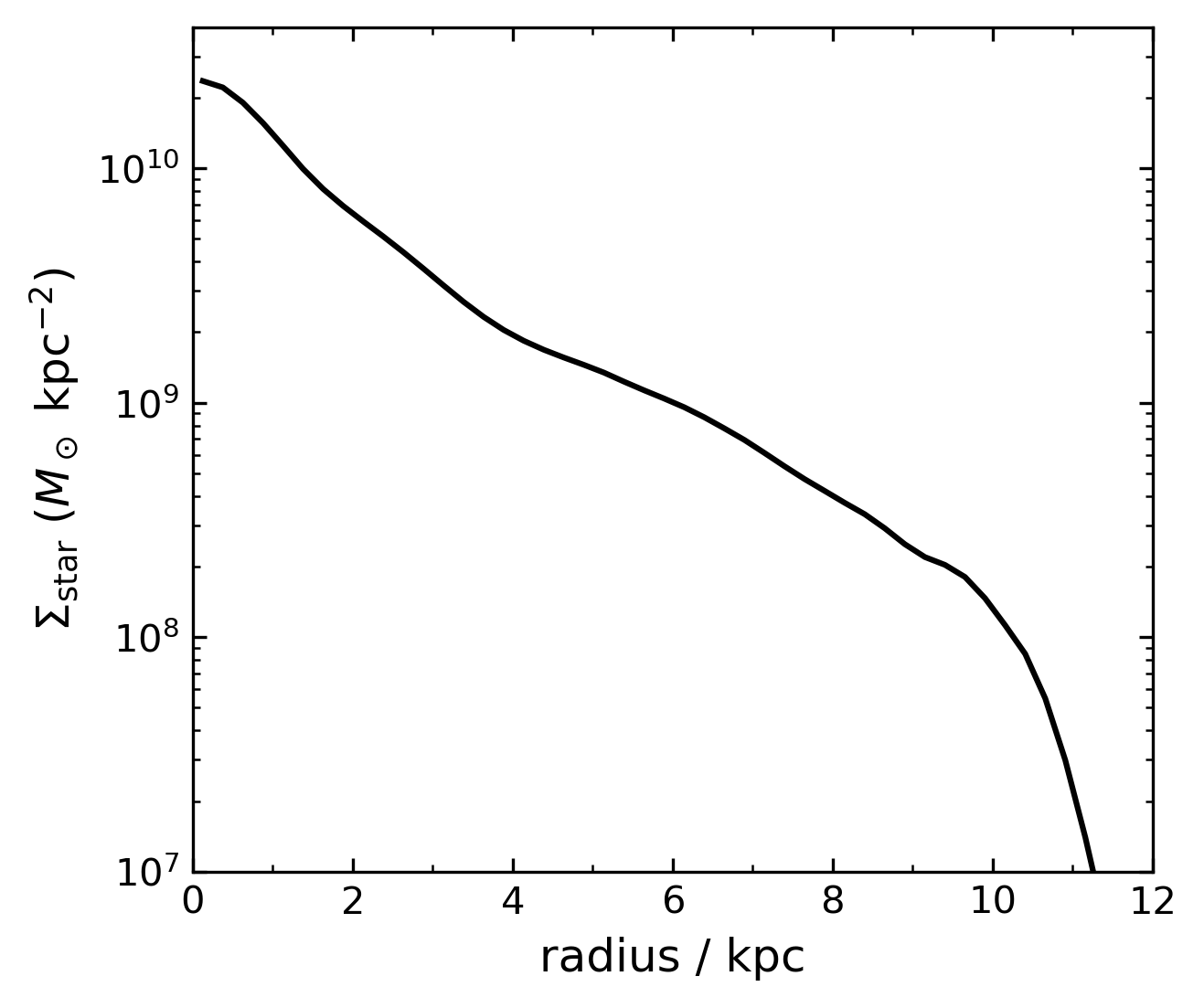}\\
    {{\bf Extended Data Figure 7 $|$ } The radial profile of stellar mass surface density in J0107a.}
    \label{fig:s7}
\end{figure}
\begin{figure}
    \center
    \includegraphics[width=0.8\linewidth]{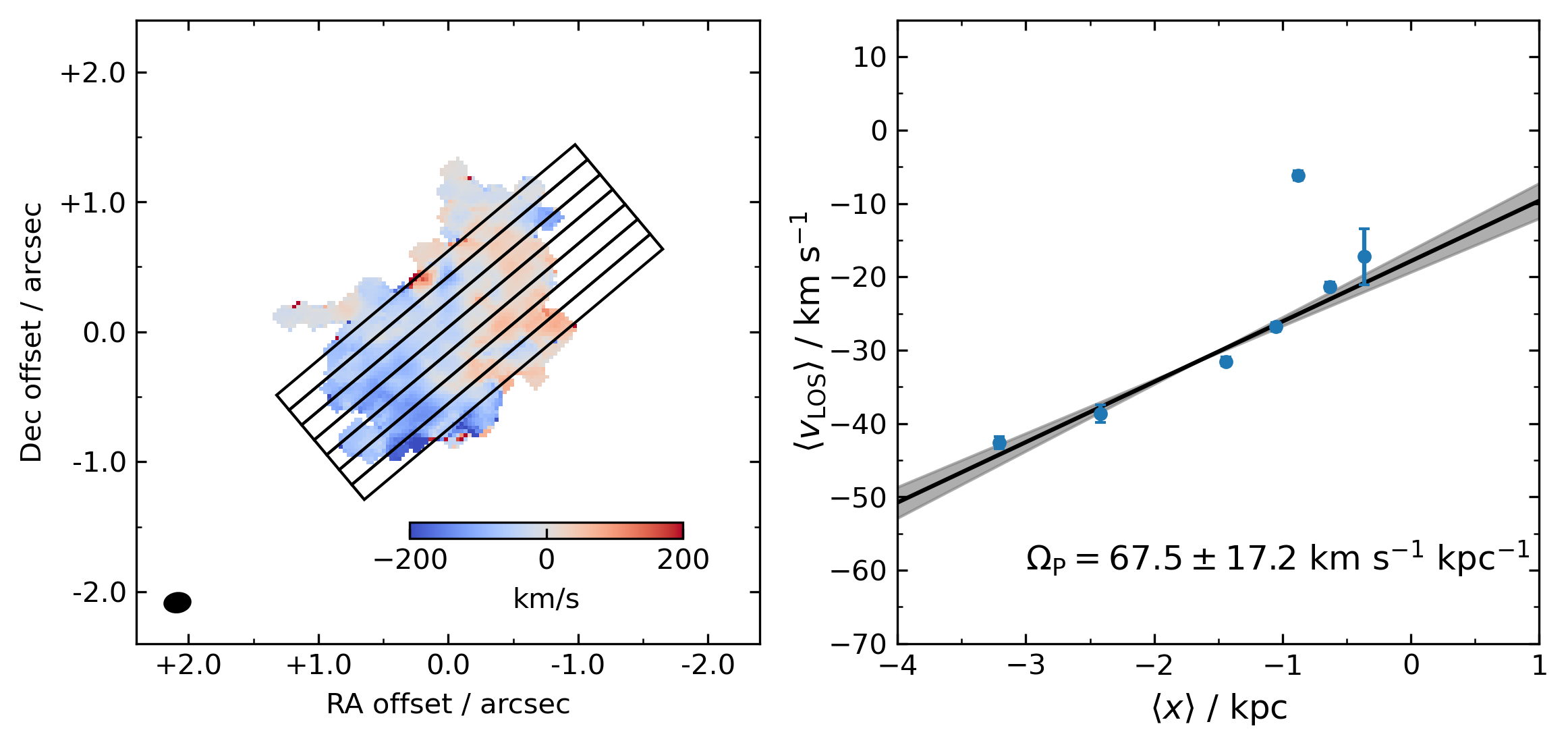}\\
    {{\bf Extended Data Figure 8 $|$ }{\bf Left}: Rectangular apertures overplot on the CO(4\textendash3) LOS velocity map; {\bf right}: the measured $\left<x\right>$-$\left<v_{\rm LOS}\right>$ relation. The best-fit linear function and uncertainty are shown as the black line and gray shaded region, respectively.}
    \label{fig:s8}
\end{figure}
\subsection{The pattern speed of J0107a's bar}
We measure the bar pattern speed with the Tremaine-Weinberg method\cite{1984ApJ...282L...5T}.  Since measuring stellar velocity field of a dusty star-forming galaxy like J0107a is impractical, we use the CO(4\textendash3) velocity map. Studies of local galaxies have found the use of ISM tracers like CO or H$\alpha$ may lead to overestimate in pattern speed up to $\sim40\%$ due to their clumpy nature\cite{2021AJ....161..185W}, but  we can expect a better performance of CO-based pattern speed measurement for J0107a because it has a $>10\times$ higher average molecular gas surface density of $1800\pm300 M_\odot$ pc$^{-2}$ within 13 kpc radius, and hence larger filling factor and less clumpy gas distribution. In addition, at such high surface density, the gas should be almost completely molecular\cite{2024A&A...691A.163E}. Therefore, CO can trace the total gas content as suggested by its similarity with [CI](1\textendash0) in distribution, and total gas content is an acceptable probe to pattern speed\cite{2023MNRAS.524.3437B}.
The pattern speed is expressed as
\begin{equation}
\Omega_{\rm P} = \frac{\left<v_{\rm LOS}\right>}{\left<x\right>\sin i},
\end{equation}
where $\left<v_{\rm LOS}\right>$ and $\left<x\right>$ are the intensity-weighted line-of-sight velocity and coordinate offset relative to the galactic center in the major axis direction, respectively. By measuring these two quantities in seven $0\farcs15\times3\farcs0$ rectangular apertures parallel to the major axis (Extended Data Figure 8, the left panel), the pattern speed can be determined as the slope of the $\left<x\right>-\left<v_{\rm LOS}\right>$ relation. We then fit a linear function to the data using orthogonal distance regression and derive $\Omega_{\rm P}=67.5\pm17.2$ km s$^{-1}$ kpc$^{-1}$ (Extended Data Figure 8, the right panel), where the error includes inclination uncertainty.
\subsection{Summary of physical properties}
\begin{table}
    \centering
    \begin{tabular}{lccc}
    \hline\hline
        Property & value & unit &note\\ \hline
        right ascension& $01^{\rm h}07^{\rm m}48\farcss3$ &\\
        declination &$-17\arcdeg30\arcmin28\farcs0$&\\
        redshift & 2.4669 &\\
        morphological type &SBbc\\
        inclination &$7\pm1$&$\arcdeg$\\
        position angle &310&$\arcdeg$\\
        stellar mass $M_\star$&$4.5^{+1.7}_{-1.4}$&$10^{11}M_\odot$&(a)\\
        star formation rate &$499^{+357}_{-179}$&$M_\odot$ ${\rm yr}^{-1}$&(a)\\
        CO(1\textendash0) luminosity &$8.1\pm1.1$&$10^{10}$ K km s$^{-1}$ pc$^{2}$\\
        molecular gas mass $M_{\rm mol}$ &$3.3\pm0.6$&$10^{11}M_\odot$&(b)\\
        gas fraction $M_{\rm mol}/M_\star$ &$0.69_{-0.29}^{+0.42}$&\\
        rotation velocity &$640\pm100$&km s$^{-1}$\\
        dynamical mass&$9\pm3$&$10^{11}M_\odot$&(c)\\
        net gas inflow rate&$579_{-174}^{+188}$&$M_\odot$ ${\rm yr}^{-1}$&(d)\\
        bar pattern speed &$67.5\pm17.2$&km s$^{-1}$ kpc$^{-1}$\\
        bar strength &0.8& &(e)\\
        bar semi-major axis length &7.5& kpc\\
        \hline
    \end{tabular}
    \caption{Summary of physical properties of J0107a. Notes: (a) from SED fitting\cite{2023ApJ...958L..26H}; (b) calculated from CO(1\textendash0) luminosity; (c) computed within 2.2 times effective radius ($\approx10$ kpc); (d) integrated within radius of 13 kpc. (e) maximum $A_2$.}
    \label{table}
\end{table}

\newpage
\textbf{References}

\bibliography{main}
\bibliographystyle{naturemag}

\end{document}